%% file: main.tex
\documentclass[reprint,preprintnumbers,amsmath,amssymb,aps,nofootinbib,showkeys,superscriptaddress]{revtex4-2}
\input{commands.tex} 

\begin{document}

\title{Tunable Mpemba Effect in a Prethermal Many-Body Spin Network}

\author{Chaitali Shah}
\thanks{These authors contributed equally to this work.}
\affiliation{Department of Chemistry, University of California, Berkeley, Berkeley, CA 94720, USA}

\author{Cooper M. Selco}
\thanks{These authors contributed equally to this work.}
\affiliation{Department of Chemistry, University of California, Berkeley, Berkeley, CA 94720, USA}

\author{Leo Joon Il Moon}
\thanks{These authors contributed equally to this work.}
\affiliation{Department of Chemistry, University of California, Berkeley, Berkeley, CA 94720, USA}

\author{Ashok Ajoy}
\email{ashokaj@berkeley.edu}
\affiliation{Department of Chemistry, University of California, Berkeley, Berkeley, CA 94720, USA}
\affiliation{Chemical Sciences Division, Lawrence Berkeley National Laboratory, Berkeley, CA 94720, USA}

\begin{abstract}
Relaxation in an interacting system is determined not only by its initial distance from equilibrium, but also by the relaxation modes populated by the initial state. Here we experimentally observe and control the Mpemba effect, in which a state farther from equilibrium overtakes one initially closer, in an extended, disordered $^{13}$C nuclear-spin network in diamond. Field cycling allows us to prepare distinct spatial polarization profiles by independently controlling hyperpolarization and defect-mediated relaxation. We then track their evolution under Floquet driving, which stabilizes a long-lived prethermal regime. We observe reproducible Mpemba crossings and tune the crossing time over several orders of magnitude, from late-time thermalization into the prethermal plateau. Semiclassical simulations show that randomly positioned paramagnetic defects create fast-relaxing regions and defect-poor regions that support the slowest collective relaxation mode. The Mpemba crossings are set by the initial state overlap with this mode. Our results demonstrate anomalous relaxation within a prethermal many-body regime and identify disorder, transport, and mode-selective state preparation as resources for controlling relaxation in extended spin networks.
\end{abstract}

\maketitle


\pagebreak

\emph{Introduction---}The relaxation of a many-body system is not determined by its initial distance from equilibrium alone, but also by how the initial state overlaps with the dynamical modes that carry it toward equilibrium. Resolving and controlling this mode structure is central to nonequilibrium quantum science~\cite{polkovnikovColloquiumNonequilibriumDynamics2011, okaFloquetEngineeringQuantum2019b, horodeckiFundamentalLimitationsQuantum2013}, with applications ranging from preserving initialized states for quantum sensing~\cite{degenQuantumSensing2017, abobeihAtomicscaleImaging27nuclearspin2019a, sahinHighFieldMagnetometry2022e} to accelerating equilibration for rapid qubit reset~\cite{baoAcceleratingQuantumRelaxation2025, lejeuneAcceleratingQubitReset2026}. The Mpemba effect provides a direct manifestation of this principle: a state initially farther from equilibrium can relax faster than one initially closer to it, causing their relaxation trajectories to cross.

Originally discussed in the context of water freezing~\cite{mpembaCool1969,burridgeQuestioningMpembaEffect2016}, the classical Mpemba effect has since been observed across a range of nonequilibrium systems~\cite{huConformationDirectedMpemba2018,ahnExperimentalVerificationsMpembalike2016,chaddahOvertakingApproachingEquilibrium2010,kumarExponentiallyFasterCooling2020,kumarAnomalousHeatingColloidal2022,tianExperimentalStudyMpemba2025}. A mode-based formulation provides a unifying picture: relaxation is anomalously fast when the initial state has reduced overlap with the slowest-decaying mode~\cite{luNonequilibriumThermodynamicsMarkovian2017,klichMpembaIndexAnomalous2019}. This picture has motivated broad extensions to open and isolated quantum systems~\cite{aresQuantumMpembaEffects2025, carolloExponentiallyAcceleratedApproach2021,navaMpembaEffectsOpen2024,strachanNonMarkovianQuantumMpemba2025,rylandsMicroscopicOriginQuantum2024,yamashikaQuantumMpembaEffect2026,yuQuantumMpembaEffects2025,summerResourceTheoreticalUnificationMpemba2026} and recent observations in trapped ions~\cite{zhangObservationQuantumStrong2025, joshiObservingQuantumMpemba2024, aharonyshapiraInverseMpembaEffect2024}, superconducting processors~\cite{xuObservationModulationQuantum2026}, and nuclear spins~\cite{chatterjeeDirectExperimentalObservation2025,schnepperExperimentalObservationApplication2025} (see Supplemental Material Sec. \ref{section_context}). However, the Mpemba effect has not yet been experimentally observed in an extended interacting network, where the required mode selectivity must emerge collectively from many-body dynamics. A recent theoretical study demonstrated that the inverse Mpemba effect can arise during relaxation toward a prethermal state in a periodically driven isolated system~\cite{sugimotoPrethermalInverseMpemba2025}; however, prethermal Mpemba dynamics have not yet been observed experimentally, and extending this phenomenon to the long-lived prethermal regime, after the system has entered the prethermal state, remains unexplored.

Here we address these questions in a disordered $^{13}$C nuclear-spin network in nitrogen-doped diamond. Optically pumped nitrogen-vacancy (NV) centers act as local polarization sources, while randomly positioned NV and substitutional nitrogen (P1) centers produce a heterogeneous landscape of relaxation sinks~\cite{beatrezElectronInducedNanoscale2023e,loweNuclearSpinLatticeRelaxation1968b,ajoyHyperpolarizedRelaxometryBased2019b}. Dipolar spin transport couples nearby nuclear spins, causing relaxation to occur through collective modes involving clusters of spins rather than independent single-spin relaxation pathways. Using magnetic-field cycling, we control polarization injection, redistribution, and depletion to prepare states with distinct spatial profiles~\cite{ajoyWideDynamicRange2019,ajoyRoomTemperatureOptical2020}. We track their subsequent evolution quasi-continuously under Floquet driving, which stabilizes a long-lived prethermal regime~\cite{beatrezFloquetPrethermalizationLifetime2021d,bukovUniversalHighfrequencyBehavior2015a,eckardtHighfrequencyApproximationPeriodically2015c}. We observe reproducible Mpemba crossings and tune the crossing time over several orders of magnitude, from the late-time thermalizing regime into the prethermal plateau. Semiclassical simulations reproduce the dynamics and identify the overlap of each initial profile with a slow mode supported by defect-poor regions as the origin of the crossings. To our knowledge, this is the first observation of the Mpemba effect in an extended many-body spin network and the first experimental demonstration of Mpemba dynamics in a prethermal state.

\begin{figure*}[htbp]
    \centering
    \includegraphics[width=1\textwidth]{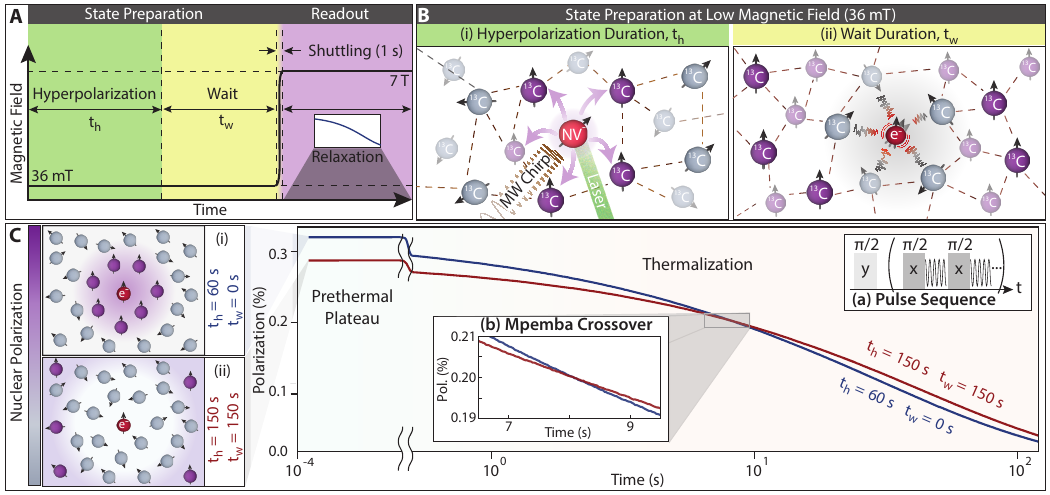}
    \caption{(A) State preparation at \val{36}{mT} consists of hyperpolarization for duration $t_{h}$ (green) followed by waiting duration $t_{w}$ (yellow). The sample is then shuttled to high field (\val{7.3}{T}) in $\approx1$ s, where spin control and readout are performed (purple). (B) (i) During $t_h$, NV centers are optically pumped and chirped microwaves transfer polarization to nearby $^{13}$C nuclei (grey, unpolarized; purple, polarized); the injected polarization spreads outward via spin diffusion. (ii) During $t_w$, free evolution leads to electron-induced relaxation of nearby $^{13}$C. (C) Left insets show two initial states: (i) $(t_h,t_w)=(60, 0) \ \mathrm{s}$ and (ii) $(t_h,t_w)=(150, 150) \ \mathrm{s}$. State (i) has larger total polarization concentrated near electrons, whereas polarization in state (ii) is distributed farther from electrons. Inset (a) shows the Floquet pulse sequence used to control and inductively measure $^{13}$C spins, producing a prethermal plateau followed by thermalization. Despite being initially farther from equilibrium, state (i) relaxes faster than state (ii), with a crossing at $t_{\times}\approx\val{8}{s}$ (inset (b)).}
    \label{}
    \zfl{fig1}
\end{figure*}

\emph{Experimental platform---}Experiments are conducted on a single-crystal diamond containing natural-abundance $^{13}\mathrm{C}$ nuclei (1.1\%), NV centers ($\approx$1 ppm), and P1 centers ($\approx$\val{30}{ppm}) (see Supplemental Material Sec. \ref{section_methods}). The $^{13}\mathrm{C}$ nuclei form an interacting spin network, coupling to one another through dipolar interactions $H_{\mathrm{nn}}=\sum_{i<j} d_{ij}\left(3I_i^z I_j^z-\vec{I}_i\cdot\vec{I}_j\right)$, with coupling strengths $d_{ij}\propto \gamma_{n}^{2}(3\cos^2(\vartheta_{ij})-1)r_{ij}^{-3}$, where $\gamma_{n}=\val{10.7}{MHz/T}$ is the gyromagnetic ratio, $r_{ij}$ is the internuclear distance, and $\vartheta_{ij}$ is the angle between the internuclear vector and the magnetic field. This secular approximation is valid since the $^{13}$C Larmor frequency is much larger than the average nearest-neighbor coupling ($\sim$\val{60}{Hz}) across all relevant magnetic fields.

Additionally, the nuclei couple to the paramagnetic defects through hyperfine interactions. Nuclei sufficiently close to electron spins experience strong Fermi contact shifts~\cite{galiInitioSupercellCalculations2008a,cox13C14N15N1994,smeltzer13CHyperfineInteractions2011,peakerAssignment13CHyperfine2016}, and are effectively excluded from bulk dynamics. The remaining spins are coupled predominantly through dipolar interactions, giving rise to relaxation rates that scale as $r^{-6}$ in both field regimes~\cite{loweNuclearSpinLatticeRelaxation1968b,reynhardtSpinLatticeRelaxation2003,beatrezElectronInducedNanoscale2023e,ajoyHyperpolarizedRelaxometryBased2019b} (see Supplemental Material Secs. \ref{subsection_lowfield} and \ref{subsection_highfield}). Owing to the sparse and random spatial distribution of the paramagnetic defects, nearby nuclei experience substantially faster relaxation than those farther away, resulting in a heterogeneous relaxation landscape.

\emph{State preparation and Mpemba crossing---}Initial states are prepared using a two-stage protocol (\zfr{fig1}A). First, at a low magnetic field of \val{36}{mT}, optically pumped NV centers inject polarization into nearby $^{13}\mathrm{C}$ nuclei over a time $t_h$ (\zfr{fig1}B(i)), following a previously described hyperpolarization mechanism~\cite{ajoyOrientationindependentRoomTemperature2018d, sarkarRapidlyEnhancedSpinPolarization2022c}. Simultaneously, dipolar spin transport redistributes this polarization throughout the nuclear-spin network. After polarization injection, the system remains at low field for an additional period $t_w$, during which fluctuating electron spins drive spatially inhomogeneous $T_1$ relaxation that selectively depletes polarization stored near both NV and P1 centers (\zfr{fig1}B(ii)).

The subsequent relaxation dynamics are probed by rapidly shuttling the sample to a high magnetic field of $B_0=\val{7.3}{T}$ in $\sim\val{1}{s}$ (\zfr{fig1}A), using a field-cycling platform developed for optical hyperpolarization experiments~\cite{ajoyWideDynamicRange2019,ajoyRoomTemperatureOptical2020}. The nuclear magnetization is then monitored using a previously described Floquet pulse sequence and high-speed NMR spectrometer~\cite{beatrezFloquetPrethermalizationLifetime2021d,moonHighspeedHighmemoryNMR2025a} (inset (a), \zfr{fig1}C). The sequence stabilizes a prethermal regime that extends the nuclear-spin lifetime and allows each relaxation trajectory to be tracked quasi-continuously over several minutes. After preparation, all states are subjected to the same Floquet sequence and relax toward the same unpolarized, effectively infinite-temperature state. For the positive-polarization states considered here, total nuclear polarization provides a valid measure of distance from equilibrium, satisfying the monotonicity, temperature-ordering, and convexity criteria~\cite{luNonequilibriumThermodynamicsMarkovian2017}.
 
The two preparation parameters $t_h$ and $t_w$ jointly control the total polarization and its spatial distribution relative to the paramagnetic defects. States with comparable polarization can therefore exhibit different relaxation dynamics. \zfr{fig1}C shows a representative example, comparing two experimentally measured relaxation trajectories prepared using $(t_h,t_w)=\val{(60, 0)}{s}$ and $\val{(150, 150)}{s}$. Although the state prepared with $(t_h,t_w)=\val{(60, 0)}{s}$ begins farther from equilibrium, it relaxes more rapidly and is eventually overtaken by the other, resulting in a clear Mpemba crossing at $t_{\times}\approx8$ s (inset (b), \zfr{fig1}C). As illustrated schematically in \zfr{fig1}C(i) and \zfr{fig1}C(ii), the faster-relaxing state stores a larger fraction of its polarization near the paramagnetic defects, where relaxation is strongest.

\begin{figure}[htbp]    
\includegraphics{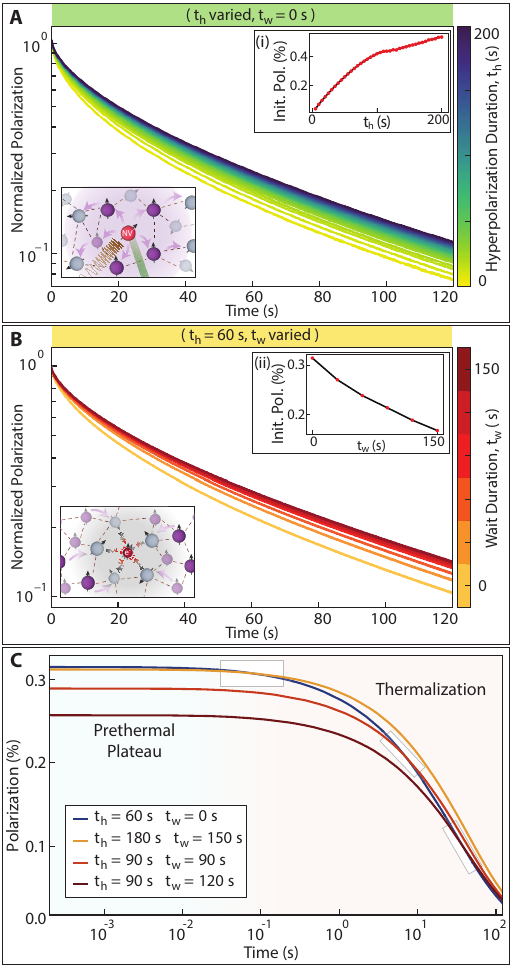}
    \caption{(A) Normalized relaxation curves for increasing $t_{h}$ with $t_{w}=\val{0}{s}$. As $t_{h}$ increases, polarization spreads farther from electrons (shown schematically in lower-left inset), reducing overlap with rapidly relaxing regions and thereby increasing lifetime. Upper-right inset shows the corresponding increase in initial polarization. (B) Normalized relaxation curves for increasing $t_{w}$ with $t_{h}=\val{60}{s}$ fixed. As $t_{w}$ increases, polarization near electrons is depleted (shown schematically in lower-left inset), again reducing overlap with rapidly relaxing regions and thereby increasing lifetime. Upper-right inset shows the corresponding decrease in initial polarization. (C) Mpemba effect between one state initially farther from equilibrium $(t_h,t_w)=(60, 0) \ \mathrm{s}$ and three other initial states (see legend). The state-preparation protocol provides tunable control over the Mpemba crossing time, shifting the curve intersections.}
    \label{}
    \zfl{fig2}
\end{figure}

\emph{Tunable Mpemba crossings---}To systematically characterize how the preparation protocol shapes the subsequent relaxation dynamics, we independently vary the hyperpolarization duration $t_h$ and the wait duration $t_w$. To minimize systematic errors, parameter values were randomized between runs. In \zfr{fig2}A, we vary $t_h$ while fixing $t_w=0$. Increasing $t_h$ produces a larger initial signal, as quantified in the upper-right inset of \zfr{fig2}A, and leads to progressively slower relaxation (normalized curves, \zfr{fig2}A). This behavior arises because polarization is generated locally near the NV centers~\cite{ajoyOrientationindependentRoomTemperature2018d,sarkarRapidlyEnhancedSpinPolarization2022c,zangaraDynamicsFrequencysweptNuclear2019a} and subsequently redistributed throughout the nuclear-spin network by dipolar spin diffusion~\cite{slichterPrinciplesMagneticResonance1990,abragamPrinciplesNuclearMagnetism1983a} (lower-left inset, \zfr{fig2}A).

In \zfr{fig2}B, we instead fix the hyperpolarization duration ($t_h=60\space\mathrm{s}$) and vary the subsequent wait duration $t_w$. Increasing $t_w$ decreases the initial signal, as quantified in the upper-right inset of \zfr{fig2}B, while simultaneously slowing the relaxation dynamics (normalized curves, \zfr{fig2}B). During this waiting period, dipolar spin diffusion continues to redistribute polarization throughout the network, while fluctuating NV and P1 electron spins drive spatially inhomogeneous nuclear-spin relaxation~\cite{loweNuclearSpinLatticeRelaxation1968b,reynhardtSpinLatticeRelaxation2003,beatrezElectronInducedNanoscale2023e,ajoyHyperpolarizedRelaxometryBased2019b} (lower-left inset, \zfr{fig2}B). Because nuclei nearest to paramagnetic defects relax most rapidly, increasing $t_w$ preferentially removes polarization stored in these regions.

Together, $t_h$ and $t_w$ provide two independent preparation knobs that jointly determine the total polarization and its spatial distribution within the nuclear-spin network. Longer hyperpolarization times increase the amount of stored polarization, whereas longer wait times selectively deplete polarization near the paramagnetic defects. Different combinations of $t_h$ and $t_w$ can therefore produce states with substantially different relaxation dynamics. This control over the initial state allows us to engineer pairs of states that exhibit the Mpemba effect and to tune the time at which their relaxation curves intersect. Representative examples are shown in \zfr{fig2}C, where varying the preparation protocol systematically shifts $t_\times$ over several orders of magnitude. The crossing can be tuned from the late-time thermalizing regime into the prethermal plateau.

\begin{figure*}[htbp!]
    \centering
    \includegraphics[width=1\textwidth]{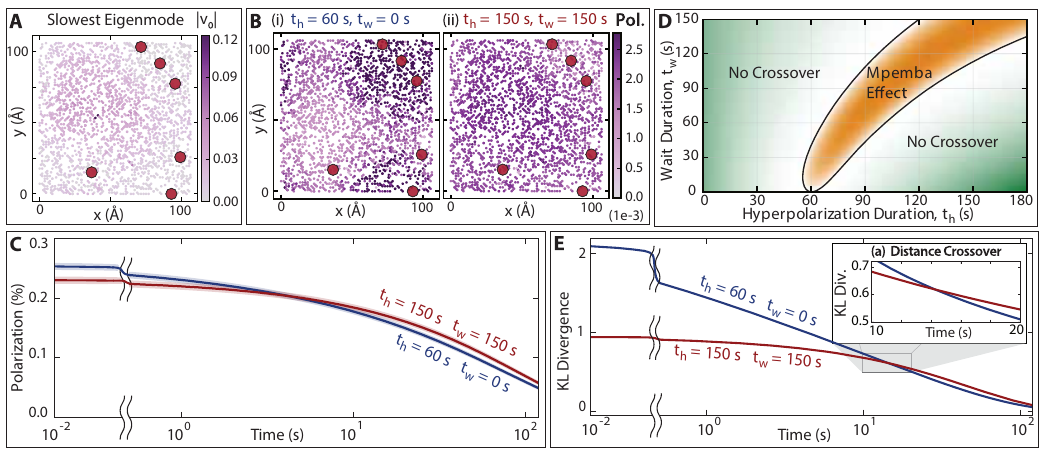}
    \caption{(A) 2D projection of spatial profile of slowest eigenmode $|v_0|$ for single lattice configuration. Mode (colorbar, purple) is localized in regions of lattice far from paramagnetic defects (red). (B) Representative initial polarization distributions for same lattice configuration, prepared with (i) $(t_h,t_w)=\val{(60,0)}{s}$ and (ii) $(t_h,t_w)=\val{(150, 150)}{s}$. State (i) carries more polarization, but less overlap with $v_{0}$ ($a_0= 1.47\times10^{-3}$); state (ii) has larger overlap with $v_0$ ($a_0= 1.71\times10^{-3}$) visible as greater density of dark (highly polarized) sites away from defects. Color scale is clipped for visualization.
(C) Configurationally averaged relaxation dynamics for the two states in (B), showing a Mpemba crossing.
(D) Phase diagram in $(t_h,t_w)$ parameter space showing regions where Mpemba crossings are predicted relative to reference state $(t_h,t_w)=(60, 0) \ \mathrm{s}$. In orange region Mpemba crossing occurs; in green regions no crossing occurs.
(E) Kullback--Leibler (KL) divergence between the polarization distribution and equilibrium, computed within the semiclassical product-state approximation and averaged over all configurations.}
    \zfl{fig3}
\end{figure*}

\emph{Relaxation-mode origin---}To elucidate the microscopic origin of the observed Mpemba effect, we employ a semiclassical polarization-transport framework~\cite{selcoEmergentDecoherenceDynamics2025b} (see Supplemental Material Sec. \ref{section_model}). For a given realization of the disordered nuclear and electron spin networks, the nuclear polarization vector  $p$ evolves according to
\begin{equation}
\dot{p}(t)=M(t)\,p(t) + I(t),
\end{equation}
where the configuration-dependent generator $M(t)$ incorporates dipolar spin transport and defect-mediated nuclear-spin relaxation~\cite{loweNuclearSpinLatticeRelaxation1968b,reynhardtSpinLatticeRelaxation2003,slichterPrinciplesMagneticResonance1990,abragamPrinciplesNuclearMagnetism1983a}, and the source term $I(t)$ describes polarization injection from optically pumped NV centers~\cite{ajoyOrientationindependentRoomTemperature2018d,sarkarRapidlyEnhancedSpinPolarization2022c} during the hyperpolarization stage. Both $M(t)$ and $I(t)$ are piecewise constant across the hyperpolarization, waiting, and readout stages of the experimental protocol. The positions of both the \C nuclei and paramagnetic defects are randomly sampled from the underlying lattice, and the experimentally observed polarization is obtained by averaging over many disorder realizations.

For a single disorder realization, the relaxation dynamics during the readout stage may be decomposed into eigenmodes,
\begin{equation}
p(t)=\sum_j a_j v_j e^{-\lambda_j t},
\end{equation}
where $v_j$ are the eigenmodes of $M$ during readout, the relaxation rates ${\lambda_j}$ are ordered as $0 < \lambda_0 \leq \lambda_1 \leq \cdots$, and the coefficients $a_j$ quantify the overlap of the initial state with each mode. As established in mode-based formulations of the Mpemba effect~\cite{luNonequilibriumThermodynamicsMarkovian2017,klichMpembaIndexAnomalous2019,carolloExponentiallyAcceleratedApproach2021}, the crossing is determined by the overlap with the slowest-decaying mode, corresponding to the smallest relaxation rate $\lambda_0$ (which averages to $\lambda_0 = \val{(7.75 \pm 0.16)\times10^{-3}}{{s}^{-1}}$ over 100 disorder realizations). In particular, an initially farther-from-equilibrium state will relax more rapidly when its overlap $a_0$ with this slow mode is less than that of a state initially closer to equilibrium.

Physically, the slowest relaxation mode is supported by defect-poor regions of the nuclear-spin network, consistent with the disorder-protected polarization domains identified in related spin-transport studies~\cite{selcoEmergentDecoherenceDynamics2025b}, and is shown by the 2D projection in \zfr{fig3}A. \zfr{fig3}B shows simulated polarization distributions corresponding to the same preparation conditions used for the representative experimental Mpemba crossing in \zfr{fig1}C, namely $(t_h,t_w)=\val{(60,0)}{s}$ and $\val{(150,150)}{s}$, respectively. Although the state in \zfr{fig3}B(ii) has lower total polarization, its spatial profile more closely resembles the slowest relaxation mode shown in \zfr{fig3}A and therefore exhibits a larger overlap with $v_0$. Consequently, this state retains a greater fraction of its polarization at long times despite being initially closer to equilibrium. As shown in \zfr{fig3}C, the more highly polarized state relaxes more rapidly because a larger fraction of its polarization is stored near paramagnetic defects, leading to a crossing of the relaxation curves and the emergence of the Mpemba effect.

Using this model, we systematically map the overlap structure across the $(t_h,t_w)$ parameter space and identify the regimes where Mpemba crossings are expected. The resulting phase diagram (\zfr{fig3}D) shows the region (orange) in which a reference state prepared with $\val{(60,0)}{s}$ relaxes more rapidly than a state prepared at $(t_h,t_w)$. 

Finally, we verify that the observed crossings persist when the distance from equilibrium is quantified using the Kullback-Leibler (KL) divergence, a metric commonly employed to rigorously define the Mpemba effect~\cite{luNonequilibriumThermodynamicsMarkovian2017,aresQuantumMpembaEffects2025,summerResourceTheoreticalUnificationMpemba2026}. Because the model neglects interspin correlations, the KL divergence is evaluated under a product-state approximation (see Supplemental Material Sec. \ref{section_kl}). \zfr{fig3}E shows the KL divergence for the same pair of states considered in \zfr{fig3}C. The KL-divergence curves cross at approximately the same time as the polarization curves, confirming that the observed Mpemba effect is not an artifact of using total polarization as a measure of proximity to equilibrium.

\emph{Outlook---}Our results show that native spatial disorder can reorder relaxation trajectories in an extended many-body system. Randomly positioned NV and P1 centers produce fast local relaxation and leave defect-poor regions that support slow collective modes. Spin transport couples clusters of spins within these regions, while state preparation controls the polarization stored within them. The resulting overlap with the slowest mode determines which trajectory persists at long times.

A central feature of the experiment is that the crossings can be shifted into the prethermal plateau. Prethermalization creates a metastable regime in which the hierarchy of relaxation modes can be resolved before final thermalization~\cite{beatrezFloquetPrethermalizationLifetime2021d,bukovUniversalHighfrequencyBehavior2015a,eckardtHighfrequencyApproximationPeriodically2015c,selcoBreakdownDisorderSuppressedFloquet2026a,sugimotoPrethermalInverseMpemba2025}. By varying only the state preparation, we shift crossings between the prethermal and thermalizing regimes without changing the drive or the final equilibrium state. This demonstrates that Mpemba physics can be controlled within a long-lived driven many-body state, where Floquet driving engineers the relaxation-mode structure and state preparation controls the populations of those modes.

The underlying mechanism is general. It requires spatially heterogeneous relaxation and control over the initial spatial distribution. These ingredients arise naturally in disordered magnetic materials~\cite{kamberSelfinducedSpinGlass2020}, molecular spin networks~\cite{ardavanWillSpinRelaxationTimes2007}, semiconductor spin ensembles~\cite{millington-hotzeNuclearSpinDiffusion2023a}, and wide-bandgap materials relevant to hyperpolarization and quantum sensing~\cite{christleIsolatedElectronSpins2015,klimovQuantumEntanglementAmbient2015}. The results suggest concrete design rules for controlling relaxation: defect density and geometry shape the slow modes, transport sets their spatial extent, and state preparation determines their weight. From this perspective, disorder can be viewed as a resource whose effects can be deliberately engineered to tailor the eigenmode structure and, consequently, the relaxation dynamics.

\emph{Acknowledgements---} We gratefully thank C. Bengs and W. Ng for insightful discussions. We acknowledge funding from ONR (N00014-20-1-2806), AFOSR YIP (FA9550-23-1-0106), DOE BES (DE-SC0026041), Dreyfus and Sloan Foundations, and instrumentation support from AFOSR DURIP (FA9550-22-1-0156) and NSF MRI (2320520). This material is based upon work supported by the Air Force Office of Scientific Research under award number FA9550-25-C-B010.

\bibliography{Mpemba} 
\input{SI.tex}

\clearpage


\end{document}

%% file: commands.tex
\usepackage[utf8]{inputenc}  
\usepackage{graphicx}  
\usepackage{xcolor}  
\usepackage{amsmath}  
\usepackage{amsfonts}  
\usepackage{latexsym}  
\usepackage{amssymb,bbm}  
\usepackage{bm}  
\usepackage[colorlinks=true,allcolors=blue]{hyperref}  
\usepackage{lipsum}  
\usepackage{setspace}
\usepackage{braket}
\usepackage{svg}
\usepackage{wrapfig}
\usepackage{booktabs}   
\usepackage{tabularx}   
\usepackage{array}      

\newcommand{\val}[2]{\ensuremath{{#1}\;\mathrm{#2}}}

\newcommand{\C}{$^{13}\mathrm{C} \ $}

\renewcommand{\figurename}{Fig.}

\newcommand{\beq}{\begin{equation}}
\newcommand{\eeq}{\end{equation}}
                  
\newcommand{\benum}{\begin{enumerate}}
\newcommand{\eenum}{\end{enumerate}}
                    
\newcommand{\bit}{\begin{itemize}}
\newcommand{\eit}{\end{itemize}}

\newcommand{\bea}{\begin{eqnarray}}
\newcommand{\eea}{\end{eqnarray}}

\newcommand{\bev}{\begin{verbatim}}
\newcommand{\eev}{\end{verbatim}}

\newcommand{\T}[1]{\textbf{#1}}

\newcommand{\zfl}[1]{\protect\label{fig:#1}}
\newcommand{\zfr}[1]{\figurename\,\ref{fig:#1}}

\newcommand{\ba}{\left\{ \begin{array}{lr}}
\newcommand{\ea}{\end{array}\right.}

\newcommand{\blist}[1]{
 \begin{list}{#1}
 \begin{align}
	 arrow
 \end{align}
 $\checkmark\star
  { \setlength{\itemsep}{3pt}
     \setlength{\parsep}{2pt}
     \setlength{\topsep}{3pt}
     \setlength{\partopsep}{0pt}
     \setlength{\leftmargin}{1em}
     \setlength{\labelwidth}{1em}
     \setlength{\labelsep}{0.5em} } }
\newcommand{\elist}{
  \end{list}  }

\DeclareMathSymbol{\vartheta}{\mathalpha}{letters}{"12}
\DeclareMathSymbol{\theta}{\mathalpha}{letters}{"23}
\DeclareMathSymbol{\phi}{\mathalpha}{letters}{"27}
\DeclareMathSymbol{\varphi}{\mathalpha}{letters}{"1E}

\newcommand{\bef}
{
\begin{figure}[htbp]
\centering
}

\newcommand{\eef}{\end{figure}}


%% file: SI.tex

\clearpage
\onecolumngrid



\setcounter{section}{0}
\setcounter{equation}{0}
\setcounter{figure}{0}
\setcounter{table}{0}
\renewcommand{\thesection}{S\arabic{section}}
\renewcommand{\theequation}{S\arabic{equation}}
\renewcommand{\thefigure}{S\arabic{figure}}
\renewcommand{\thetable}{S\arabic{table}}
\renewcommand{\thesubsubsection}{\thesubsection.\arabic{subsubsection}}
\setcounter{secnumdepth}{3}
\setcounter{tocdepth}{2}

\renewcommand{\tocname}{Supplementary Information}
\tableofcontents

\section{Methods}
\label{section_methods}

\T{Materials---} The sample used in this work is a (100)-cut, type-Ib, single-crystal diamond with dimensions $3.4 \times 3.2 \times 2.1$ mm. The crystal contains naturally abundant $^{13}\mathrm{C}$ nuclei (1.1\%), nitrogen-vacancy (NV) centers ($\approx 1$ ppm), and substitutional nitrogen (P1) centers ($\approx 30$ ppm). The average nearest-neighbor separation between $^{13}\mathrm{C}$ nuclei is approximately $4.5~\mathrm{\AA}$, and the average nearest-neighbor dipolar coupling strength is $\approx60$ Hz. The nuclear free-induction decay (FID) time, $T_{2}^{\star}$, is $\approx 1.5$ ms~\cite{beatrezFloquetPrethermalizationLifetime2021d}. The nuclear spin-lattice relaxation times are $T_1 \approx 283$ s at low magnetic field (36 mT) and $T_1 \approx 1520$ s at high magnetic field (7.3 T)~\cite{beatrezElectronInducedNanoscale2023e}.

\T{Experimental setup -- }
The experiments were performed using a home-built field-cycling apparatus described in previous work~\cite{ajoyWideDynamicRange2019, ajoyRoomTemperatureOptical2020, sarkarRapidlyEnhancedSpinPolarization2022c}. The diamond sample is housed in a glass sample tube and immersed in water to improve laser illumination uniformity during hyperpolarization and to facilitate thermal management. The low-field region of the apparatus consists of a Helmholtz coil that generates a magnetic field of 36 mT, a microwave excitation coil delivering approximately 15 W of power at $\approx4$ GHz, and a laser illumination assembly composed of thirty 1-W, 532-nm fiber-coupled lasers that simultaneously illuminate the sample. Together, the optical and microwave excitation drive the transfer of polarization from NV centers to nearby $^{13}$C nuclei, as described in previous works~\cite{ajoyOrientationindependentRoomTemperature2018d, zangaraDynamicsFrequencysweptNuclear2019a}.

The sample tube is attached to a belt-driven actuator that rapidly shuttles the sample between the low-field hyperpolarization region and a superconducting magnet operating at 7.3 T in $\approx1$ s. Inside the magnet, a home-built NMR probe equipped with a copper saddle coil provides coherent control and inductive detection of the $^{13}\mathrm{C}$ nuclear-spin ensemble at $\sim75$ MHz.

\(^{13}\mathrm{C}\) interrogation is carried out with a home-built NMR spectrometer built around a high-speed arbitrary waveform transceiver (Proteus P9484M)~\cite{moonHighspeedHighmemoryNMR2025a}. This AWT platform is used both to synthesize the NMR control pulses and to directly digitize the nuclear precession signal at the Larmor frequency, thereby avoiding the insertion losses that are often introduced by intermediate-frequency conversion. Its large onboard memory (16\,GB) and high sampling rate (up to 2.7\,GS/s) make it possible to continuously monitor \(^{13}\mathrm{C}\) spin precession during the intervals between pulses. The full Larmor precession is sampled at intervals of \(0.74\,\mathrm{ns}\) and mixed with an onboard numerically controlled oscillator (NCO) tuned to the Larmor frequency, enabling direct tracking of both the spin amplitude and phase in the rotating frame.

For the relaxation measurements reported in this work, the $^{13}\mathrm{C}$ nuclear-spin ensemble is first rotated into the transverse plane by a $(\pi/2)_y$ pulse and subsequently subjected to a Floquet pulse sequence consisting of a train of $(\pi/2)_x$ pulses. The pulse sequence engineers an effective nuclear dipolar Hamiltonian that creates prethermal states with greatly prolonged lifetimes, as described in a previous work~\cite{beatrezFloquetPrethermalizationLifetime2021d}. Here, the pulse duration is $56.32~\mu\mathrm{s}$, with an inter-pulse delay of $64.92~\mu\mathrm{s}$, and a typical experiment contains $\sim1$ M pulses. During each inter-pulse interval, the nuclear precession signal is recorded and digitized for $3~\mu\mathrm{s}$. The acquired time-domain signal is Fourier transformed, and the amplitude of the resulting spectral peak is tracked as a function of time to obtain the nuclear-spin relaxation trajectory.

\section{Comparison with previous experimental literature}
\label{section_context}

The Mpemba effect has been observed across a wide range of platforms, which we briefly survey to situate the present work.

\textit{Phase-transition systems}---The earliest observations involved macroscopic systems relaxing through a first-order phase transition: water~\cite{mpembaCool1969, burridgeQuestioningMpembaEffect2016}, clathrate hydrates~\cite{ahnExperimentalVerificationsMpembalike2016}, polylactide crystallization~\cite{huConformationDirectedMpemba2018}, and magnetic manganites~\cite{chaddahOvertakingApproachingEquilibrium2010}. In these systems the anomalous relaxation is tied to nucleation barriers or metastable free-energy landscapes, and it long remained unclear whether the effect was intrinsic to such metastability or generic to relaxation. Lu and Raz~\cite{luNonequilibriumThermodynamicsMarkovian2017} resolved this by recasting the Mpemba effect in a Markovian framework requiring no phase transition: a state relaxes anomalously fast when its overlap with the slowest-decaying relaxation mode is small. This eigenmode picture underlies all subsequent controlled studies, including ours.

\textit{Engineered single-particle systems}---A second class realizes the effect in a single degree of freedom in an externally imposed potential. Kumar and Bechhoefer~\cite{kumarExponentiallyFasterCooling2020, kumarAnomalousHeatingColloidal2022} observed the Mpemba, strong Mpemba, and inverse effects for a colloidal particle in a tilted double-well optical trap, providing the first quantitative test of eigenmode-overlap theory, and Tian \textit{et al.}~\cite{tianExperimentalStudyMpemba2025} extended this to a levitated nanoparticle. Both offer continuous trajectory monitoring, but through a single engineered coordinate rather than an interacting system.

\textit{Quantum platforms}---A separate line concerns the \textit{quantum} Mpemba effect, defined not through a population distribution but through a quantum measure such as entanglement asymmetry or a coherence-based free energy. It has been demonstrated in trapped ions~\cite{zhangObservationQuantumStrong2025, aharonyshapiraInverseMpembaEffect2024, joshiObservingQuantumMpemba2024}, a superconducting processor~\cite{xuObservationModulationQuantum2026}, and few-spin NMR systems~\cite{chatterjeeDirectExperimentalObservation2025, schnepperExperimentalObservationApplication2025}, in each case involving one to a few well-controlled degrees of freedom.

\textit{This work}---Our experiment differs in that the effect arises from native spatial disorder rather than an engineered potential, phase transition, or tailored dissipation. The random placement of paramagnetic defects produces a heterogeneous $r^{-6}$ relaxation landscape whose slow eigenmodes are localized in regions far from the defects, so that the prepared spatial polarization profile, not merely the total polarization, sets the overlap that governs the crossover. The Floquet-based readout then resolves each trajectory with millions of data points; we emphasize that this driving only monitors the relaxation and does not itself produce the crossover.

\section{Modelling Polarization Dynamics}
\label{section_model}

\begin{figure}[htbp]
    \centering    \includegraphics[width=0.6\linewidth]{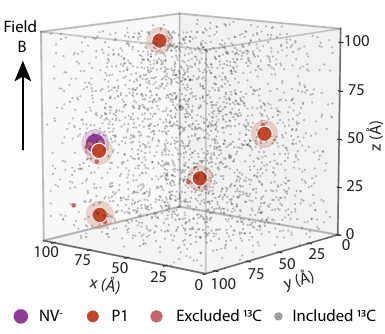}
    \caption{Representative lattice realization showing the positions of $^{13}$C nuclei (gray), P1 centers (red), and the NV center (purple). Spins within the  radius of \val{7}{\r{A}} around each paramagnetic center are excluded from the simulation.}
    \label{lattice}
\end{figure}

To model the polarization dynamics of the disordered $^{13}$C network, we employ a semiclassical approach in which polarization transport is treated as a Markovian hopping process, while relaxation and injection are modeled as on-site, spatially dependent decay and growth processes, respectively. 

We construct a diamond-cubic lattice of $30 \times 30 \times 30$ conventional unit cells, yielding a total of $\sim 2.16 \times 10^5$ lattice sites. Lattice sites are then randomly occupied according to the experimentally determined concentrations: $^{13}$C nuclei at 1.1\%, P1 centers at $\sim$30 ppm, and NV centers at $\sim$1 ppm. To ensure that each disorder realization contains at least one NV center, required for simulating the hyperpolarization process, the first defect is chosen to be an NV. This procedure yields a typical realization containing 1 NV center and 5 P1 centers. Periodic boundary conditions are imposed to suppress finite-size effects and approximate bulk dynamics.

Around each paramagnetic impurity (NV and P1), we exclude $^{13}$C nuclei within a radius of \val{7}{\r{A}}, corresponding to the region with strong Fermi contact interactions ~\cite{cox13C14N15N1994, galiInitioSupercellCalculations2008a, smeltzer13CHyperfineInteractions2011, peakerAssignment13CHyperfine2016}. After this exclusion, a typical realization contains $N_C \sim 2350$ active $^{13}$C spins. For comparison with experiment, all spins beyond the $7$ \r{A} exclusion participate in the simulated dynamics, but the reported polarization is summed only over those detectable by the readout, i.e. spins whose hyperfine coupling to every electron falls below the excitation bandwidth of the readout pulses ($\approx 17.8$ kHz). For reference, neglecting angular dependence, \val{17.8}{kHz} corresponds to a distance of $10.4$ \AA, however the full angular dependence is incorporated in the simulations. The strongly coupled spins in the intervening region are retained in the dynamics but shifted outside the detection window, so they contribute to relaxation and transport without appearing in the measured signal. We find that this system size is sufficient to capture the relevant dynamics, including the emergence of the Mpemba effect.

\textit{Governing equations}---The experimental protocol consists of three sequential stages: low-field hyperpolarization (duration $t_h$), low-field waiting (duration $t_w$), and high-field readout, spanning a total duration $T$. For a given lattice realization, the polarization vector $p(t) \in [-1,1]^{N_C}$, whose $i$-th component gives the polarization of the $i$-th $^{13}$C spin, evolves according to
\begin{equation*}
\dot{p}(t) = M(t)\,p(t) + I(t),
\label{eq:eom}
\end{equation*}
where $M(t)$ is a configuration-dependent generator incorporating spin transport and relaxation, and $I(t)$ is a source vector representing polarization injection from optically pumped NV centers. Both are piecewise constant across the three stages of the protocol:
\begin{equation*}
M(t)=
\begin{cases}
M^{\mathrm{LF}}, & 0<t<t_h+t_w \\[4pt]
M^{\mathrm{HF}}, & T>t_h+t_w
\end{cases}
\qquad
I(t)=
\begin{cases}
I, & 0<t<t_h \\[4pt]
0, & t>t_h
\end{cases}
\label{eq:stages}
\end{equation*}

\noindent The construction of each generator and the source term is detailed in the sections that follow.

For a single disorder realization, the net nuclear polarization is $p_{\mathrm{tot}}(t) = \sum_i p_i(t)$, where the sum runs over all active $^{13}$C spins. The experimentally measured polarization corresponds to a disorder-averaged quantity,
\begin{equation*}
P(t) = \langle p_{\mathrm{tot}}(t) \rangle_{\mathrm{conf}},
\end{equation*}
with good convergence typically achieved using $\sim$100 disorder realizations. Throughout the following sections, Latin indices ($i, j, \ldots$) denote nuclear spins and Greek indices ($\alpha, \beta, \ldots$) denote electronic spins.

 The present framework extends a previously developed polarization-transport model~\cite{selcoEmergentDecoherenceDynamics2025b} in several respects: we incorporate bath-induced broadening of the nuclear zero-quantum spectrum and its effect on the hopping dynamics, and we explicitly model both the polarization injection process and the low-field nuclear relaxation, including the influence of non-secular hyperfine interactions. These extensions are necessary to capture the full sequence of experimental stages considered in this work and their impact on the observed polarization dynamics.

\subsection{Low-field state preparation}
\label{subsection_lowfield}
 
During the low-field stages of the protocol ($B \approx 36$ mT), the polarization evolves as
\begin{equation*}
\dot{p}(t) = \left(W^{\mathrm{LF}} + R^{\mathrm{LF}}\right)p(t) + I(t),
\end{equation*}
where $W^{\mathrm{LF}}$ describes dipolar spin transport and $R^{\mathrm{LF}}$ electron-induced relaxation. The injection source is active only during hyperpolarization, $I(t) = I$ for $0 < t < t_h$. We now describe each term in detail.

We write the off-diagonal elements of the hopping matrix $W^{\mathrm{LF}}$, where $W^{\mathrm{LF}}_{ij}$ $(i\neq j)$ denotes the hopping rate between nuclear spins $i$ and $j$, as 
\begin{equation*}
W^{\mathrm{LF}}_{ij} = \kappa \cdot \frac{d_{ij}^{2}/2}{\tau_{n}\sum_{k\neq i,j} (d_{ik}-d_{jk})^{2} + \tau_{e}\sum_{\mu} (d_{i\mu}-d_{j\mu})^{2}}.
\end{equation*}
The diagonal elements are chosen to satisfy 
\begin{equation*}
W^{\mathrm{LF}}_{ii} = -\sum_{j\neq i}W^{\mathrm{LF}}_{ij}
\end{equation*}
which ensures conservation of the total polarization and physically corresponds to the rate at which polarization escapes from site $i$. Here, $d_{ij}$ and $d_{i\mu}$ denote the dipolar coupling constants between nuclear spin $i$ and either nuclear spin $j$ or paramagnetic impurity $\mu$ (NV or P1 center), respectively, with expressions given by:
\begin{equation*}
d_{ij}=-\frac{\mu_0}{4\pi}\frac{\hbar\gamma_C^2}{r_{ij}^3}\frac{(3{\rm cos}^{2}(\theta_{ij})-1)}{2},
\qquad
d_{i\mu}=-\frac{\mu_0}{4\pi}\frac{\hbar\gamma_C\gamma_e}{r_{i\mu}^3}\frac{(3{\rm cos}^{2}(\theta_{i\mu})-1)}{2}.
\end{equation*}
Here, $\mu_{0}$ is the permeability of free space, $\hbar$ is the reduced Planck's constant, $\gamma_{C}$ and $\gamma_{e}$ are the gyromagnetic ratios of $^{13}$C and electrons, respectively, $r_{ij}$ and $r_{i\mu}$ are the distances between nuclear spin $i$ and either nuclear spin $j$ or paramegntic impurity $\mu$, and $\theta_{ij}$ and $\theta_{i\mu}$ are the angles between the inter-spin vectors and the external magnetic field. Additionally, $\tau_{n}$ and $\tau_{e}$ denote the correlation times of the nuclear and electron baths, respectively. For a detailed derivation of $W^{\rm LF}_{ij}$, see Sec.~\ref{subsection_hopping}. Essentially, $d_{ij}^{2}$ represents the strength of the dipolar coupling between the two spins while the term in the denominator represents a broadening of the energy levels of the two spin system from fluctuations in the surrounding nuclear and electron spin baths. The dimensionless prefactor $\kappa$ is treated as an adjustable parameter that accounts for the approximations inherent in the semi-classical hopping model and is fixed to $\kappa=0.13$ throughout all simulations.

Relaxation of the $^{13}$C nuclei is approximated as arising from fluctuating fields from surrounding electron spins, resulting in a $\propto r^{-6}$, on-site decay. At low magnetic field (36 mT), one must take into account non-secular components of the dipolar Hamiltonian which are typically truncated at high field. This has been worked out previously~\cite{loweNuclearSpinLatticeRelaxation1968b}, and is given by
\begin{equation*}
R^{\rm LF}_{ij}=-\eta^{\rm LF}\delta_{ij}\frac{9}{4}\left(\frac{\mu_0}{4 \pi}\right)^2\hbar^{2}\gamma_{e}^{2}\gamma_{C}^{2}\frac{\tau_{e}}{1+\omega_n^{2}\tau_{e}^{2}}\sum_{\mu}\frac{{\rm sin^{2}}\theta_{i\mu}{\rm cos^{2}}\theta_{i\mu}}{r_{i\mu}^{6}}
\end{equation*}
where $\delta_{ij}$ is the Kronecker delta and $\omega_{n}$ is the $^{13}$C Larmor frequency. The dimensionless prefactor $\eta^{\rm LF}$ is treated as an effective model parameter and is chosen to achieve quantitative agreement with the experimentally observed low-field relaxation dynamics, yielding $\eta^{\rm LF}=0.0022$, and is fixed to this value throughout all simulations. This parameter is intended to account for the fact that the simulations employ a simplified description of the electronic environment, containing only a limited number of electron spins and neglecting several microscopic details of the true system. For example, the hyperfine interaction between P1 center electrons and their host $^{14}$N nuclei splits the electronic reservoir into three manifolds separated by $\approx114$ MHz, which is not explicitly included in the present treatment. 

We note that some previous treatments~\cite{duijvestijnQuantitativeInvestigationDynamic1986, reynhardtSpinLatticeRelaxation2003} include an additional integral term in the expression for $R^{\rm LF}_{ij}$ that quantifies the density of electron pairs separated in energy by a nuclear Larmor frequency, and therefore the availability of electronic dipolar-reservoir modes capable of mediating nuclear relaxation. However, when the EPR linewidth is much larger than the nuclear Larmor frequency, this factor approaches unity and does not modify the relaxation rate. This is the regime relevant to the present work~\cite{ajoyHyperpolarizedRelaxometryBased2019b}.

During the hyperpolarization stage, polarization is injected into the $^{13}$C network by optically pumped NV centers under simultaneous application of linearly swept microwaves. The microscopic mechanism can be understood within a two-spin model consisting of a single NV electronic transition manifold ($\ket{m_{s}=0}\rightarrow\ket{m_{s}=\pm1}$) coupled to a single $^{13}$C nucleus. As the microwave frequency is swept across the NV resonance, the system traverses a sequence of Landau-Zener level anticrossings (LZ-LACs) in the appropriate interaction-picture energy spectrum. The resulting nuclear polarization is determined by the interplay between the microwave sweep rate and the Landau-Zener transition probabilities associated with these anticrossings and has been derived previously~\cite{sarkarRapidlyEnhancedSpinPolarization2022c}. For a nuclear spin $i$, the rate of polarization buildup is given by
\begin{equation*}
I_{i} = \left[1-{\rm exp}\left(-\frac{c_{e}\eta_{e}}{f_{r}}\right)\right] f_{r} (1-\mathcal{P}_{1}) \left[1-(2\mathcal{P}_{2,i}-1)^{2}\right].
\end{equation*}
The prefactor 
\begin{equation*}
\left[1-{\rm exp}\left(-\frac{c_{e}\eta_{e}}{f_{r}}\right)\right]
\end{equation*}
represents the rate of electron polarization buildup from optical pumping, where $\eta_{e}$ is the laser power in Watts, $c_{e}$ is a proportionality constant, and $f_{r}$ is the microwave sweep repetition rate. Since a quantitative microscopic description of the optical pumping process lies beyond the scope of the present model, we absorb this factor into a phenomenological parameter, $\xi$, which we treat as an adjustable model parameter. The injection rate therefore becomes
\begin{equation*}
I_{i} = \xi f_{r} (1-\mathcal{P}_{1}) \left[1-(2\mathcal{P}_{2,i}-1)^{2}\right].
\end{equation*}
The quantities $\mathcal{P}_1$ and $\mathcal{P}_{2,i}$ are the Landau-Zener transition probabilities associated with the relevant LZ-LACs and are given by
\begin{equation*}
\mathcal{P}_1 = \exp\!\left(-\frac{\epsilon_1^2}{2\pi f_r \mathcal{B}}\right) 
\qquad \text{and} \qquad
\mathcal{P}_{2,i} = \exp\!\left(-\frac{\epsilon_{2,i}^2}{2\pi f_r \mathcal{B}}\right),
\end{equation*}
where $\mathcal{B}$ is the bandwidth over which the microwave frequency is chirped, and $\epsilon_{1}$ and $\epsilon_{2,i}$ are the energy gaps associated with the LZ-LACs,
\begin{equation*}
\epsilon_{1} = \Omega_{e}
\qquad \text{and} \qquad
\epsilon_{2} = \frac{\Omega_e \, A_{\perp,i}}{4(\omega_n + A_{\parallel,i})}.
\end{equation*}
Here, $\Omega_{e}$ is the electronic Rabi frequency, and $A_{\perp,i}$ and $A_{\parallel,i}$ are the perpendicular and parallel components of the NV-$^{13}$C hyperfine coupling for nuclear spin $i$, respectively.

The microwave sweep rate and bandwidth used in both experiments and simulations are $f_{r}=750$ Hz and $\mathcal{B} = 10$ MHz, respectively. The microwave Rabi frequency $\Omega_e$ is estimated from the experimental microwave power ($P_{\mathrm{MW}} = 15$ W) and the coil geometry, and is given by $\Omega_{e}\sim2.36$ MHz. Finally, the parameter $\xi$ is chosen such that the injected polarization in the simulation matches the experimentally inferred value. To determine the absolute polarization in the experiment, we first estimate the thermal nuclear polarization from a Boltzmann distribution, and then compare the hyperpolarized signal to the thermal signal. As shown in \zfr{thermal}, the hyperpolarized signal is enhanced by a factor of $\varepsilon = 524$ relative to the thermal signal. Multiplying this enhancement factor by the thermal polarization yields the absolute polarization corresponding to a given hyperpolarization time; for example, a hyperpolarization duration of 60 s corresponds to an absolute nuclear polarization of approximately 0.3\%. The parameter $\xi$ is then chosen so that the polarization buildup in the simulation best reproduces the experimentally observed dependence on $t_{h}$. Importantly, a single value of $\xi=10^{-5}$ is used for all simulations; it is not varied with $t_{h}$. The agreement of polarization buildup between simulation and experiment as a function of $t_h$ is shown in \zfr{buildup}. 

\begin{figure}
    \centering
    \includegraphics[width=0.5\textwidth]{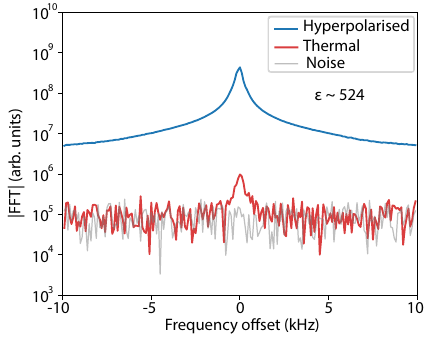}
    \caption{$^{13}$C NMR spectra showing the noise floor (gray), the thermal signal acquired at room temperature and 7.3 T (red), and the hyperpolarized signal ($T_h$ = 60 s) (blue). The hyperpolarized signal exhibits an enhancement factor of $\epsilon = 524$ relative to the thermal signal. The absolute nuclear-spin polarization is estimated by multiplying this enhancement factor by the thermal polarization calculated from a Boltzmann distribution.}
    \zfl{thermal} 
\end{figure}

\begin{figure}[htbp]
    \centering
    \includegraphics[width=0.6\linewidth]{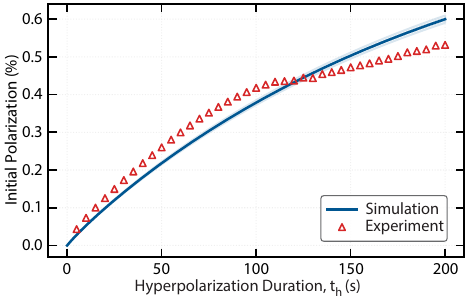}
    \caption{Polarization buildup curves as a function of hyperpolarization duration $t_{h}$ for both experiment (red triangles) and simulation (blue curve).}
    \zfl{buildup}
\end{figure}

After the hyperpolarization stage, polarization injection is stopped by turning off the laser and microwaves, and the system freely evolves in the low magnetic field. Therefore, the evolution during the waiting time is simply governed by
\begin{equation*}
\dot{p}(t)=(W^{\rm LF}+R^{\rm LF})p(t).
\end{equation*}

\begin{figure}[htbp]
    \centering
    \includegraphics{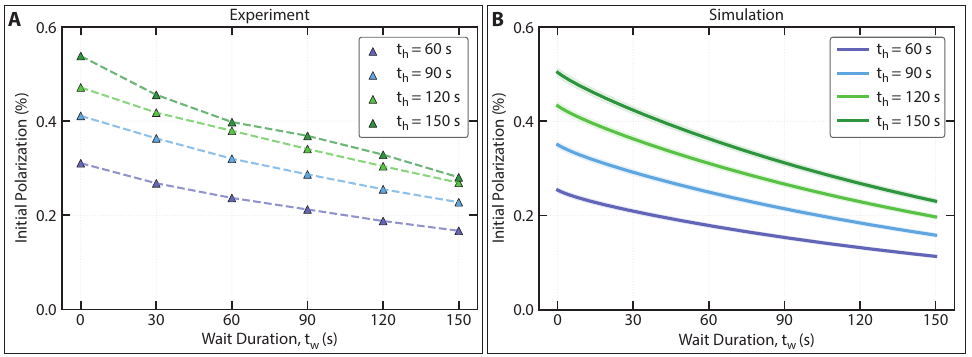}
    \caption{Initial $^{13}$C polarization as a function of wait duration $t_w$ for several hyperpolarization times $t_h$, showing the depletion of polarization during the low-field wait in both (A) experiment and (B) simulation.}
    \zfl{wait} 
\end{figure}

The simulated low-field waiting dynamics show excellent agreement with the experimentally measured polarization decay across all hyperpolarization durations. \zfr{wait}(A) shows the experimentally measured \emph{initial polarization} as a function of wait duration $t_w$ for several hyperpolarization times $t_h$, while \zfr{wait}(B) shows the corresponding simulated dynamics. The model captures not only the magnitude of the initial polarization after preparation, but also its subsequent depletion during the low-field wait. The agreement between experiment and simulation provides validation of the full semiclassical framework, including the interplay between spatially dependent polarization injection, dipolar transport, and electron-mediated relaxation.

\subsection{High Field Readout : Floquet driving}
\label{subsection_highfield}
After rapidly shuttling the sample to a high magnetic field of 7.3 T, the spins are rotated onto the x-axis in the rotating frame and a train of equally spaced x-pulses are applied. Within the semi-classical approach, the polarization dynamics during the Floquet driving are expressed as 
\begin{equation*}
\dot{p}(t)=(W^{\rm HF}+R^{\rm HF})p(t),
\end{equation*}
where $W^{\rm HF}$ and $R^{\rm HF}$ are matrices responsible for polarization transport and electron-induced relaxation at high field, respectively. 

Diffusive processes are described similarly to low field as follows
\begin{equation*}
W^{\rm HF}_{ij} = \frac{1}{4}W^{\rm LF}_{ij}
\end{equation*}
where the $1/4$ arises from a driving-protocol-dependent scaling factor. More specifically, the Floquet driving engineers an effective first-order average nuclear dipolar Hamiltonian given by 
\begin{equation*}
\bar{H}_{dd}^{(1)}=-\frac{1}{2}\sum_{i<j}d_{ij}(3I_{ix}I_{jx}- \vec{I_i} \cdot \vec{I_j}),
\end{equation*}
so the nuclear dipolar couplings are effectively scaled by $1/2$ and transition rates by $1/4$.

Relaxation of the $^{13}$C nuclei at high field is expressed as 
\begin{equation*}
R^{\rm HF}_{ij}=-\eta^{\rm HF}\delta_{ij}J_{\rm filtered}(\omega_{\rm eff})\sum_{\mu}d_{i\mu}^{2},
\end{equation*}
where $\eta^{\rm HF}$ is a phenomenological model parameter, analogous to that introduced previously, which accounts for the minimal microscopic description of the true system. It is taken to be a fixed value of $\eta^{\rm HF}=0.005$ for all simulations. Additionally, $J_{\rm filtered}(\omega)$ is the longitudinal noise spectral density of the electrons filtered by the Floquet driving protocol, and $\omega_{\rm eff}$ is the effective rotation frequency of the nuclear spins over one pulse and one inter-pulse delay of the Floquet driving sequence. The filtered spectral density is given by 
\begin{equation*}
J_{\rm filtered}(\omega)=\int_{-\infty}^{+\infty}J_{\rm bare}(\omega-\omega')Y(\omega')d\omega'
\end{equation*}
which represents a convolution between the bare longitudinal electron spectral density $J_{\rm bare}(\omega)$ and the driving-protocol-dependent spectral filter function, as shown schematically in \zfr{spectral_overlap}, and is closely analogous to the noise-filtering framework commonly used to describe dynamical decoupling sequences~\cite{cywinskiHowEnhanceDephasing2008a, uysOptimizedNoiseFiltration2009a, ajoyOptimalPulseSpacing2011a, alvarezMeasuringSpectrumColored2011b}. $J_{\rm bare}(\omega)$ is assumed to be Lorentzian, 
\begin{equation*}
J_{\rm bare}(\omega)=\frac{\tau_{e}}{1+\tau_{e}^{2}\omega^{2}},
\end{equation*}
while $Y(\omega)$ is given by 
\begin{equation*}
Y(\omega)=\sum_{k=-\infty}^{+\infty}\vert c_{k}\vert^{2}\delta(k\omega_{d}+\omega).
\end{equation*}
Here, $\omega_{d}$ is the driving frequency of the Floquet driving protocol defined as $\omega_{d}=\frac{2\pi}{T}$, where $T$ is the length of one pulse plus one inter-pulse delay. Additionally, $c_{k}$ are the Fourier coefficients, given by 
\begin{equation*}
c_{k}=T^{-1}\sum_{m=-1}^{+1}e^{+i m \phi_{\rm eff}}d^{1}_{1m}(\theta_{\rm eff})\int_{0}^{T}D^{1}_{m0}{[}P^{\dagger}(t){]}e^{-i k \omega_{d}}dt.
\end{equation*}
Here, $\theta_{\rm eff}$ and $\phi_{\rm eff}$ parameterize the effective quantization axis, $d^{l}_{mn}$ are reduced Wigner matrix elements, and $D^{l}_{m0}{[}P^{\dagger}(t){]}$ are full Wigner matrix elements parametrized by the time-periodic micromotion operator $P^{\dagger}(t)$ as defined within the Floquet framework. The Fourier coefficients can be obtained analytically using a bimodal high-frequency expansion of the Floquet Hamiltonian~\cite{scholzOperatorbasedFloquetTheory2010b, bukovUniversalHighfrequencyBehavior2015a, eckardtHighfrequencyApproximationPeriodically2015c, ivanovFloquetTheoryMagnetic2021a}. We refer the interested reader to Ref.~\cite{selcoBreakdownDisorderSuppressedFloquet2026a} for the details of the full derivation. 

\begin{figure}[htbp]
    \centering
    \includegraphics[width=0.6\linewidth]{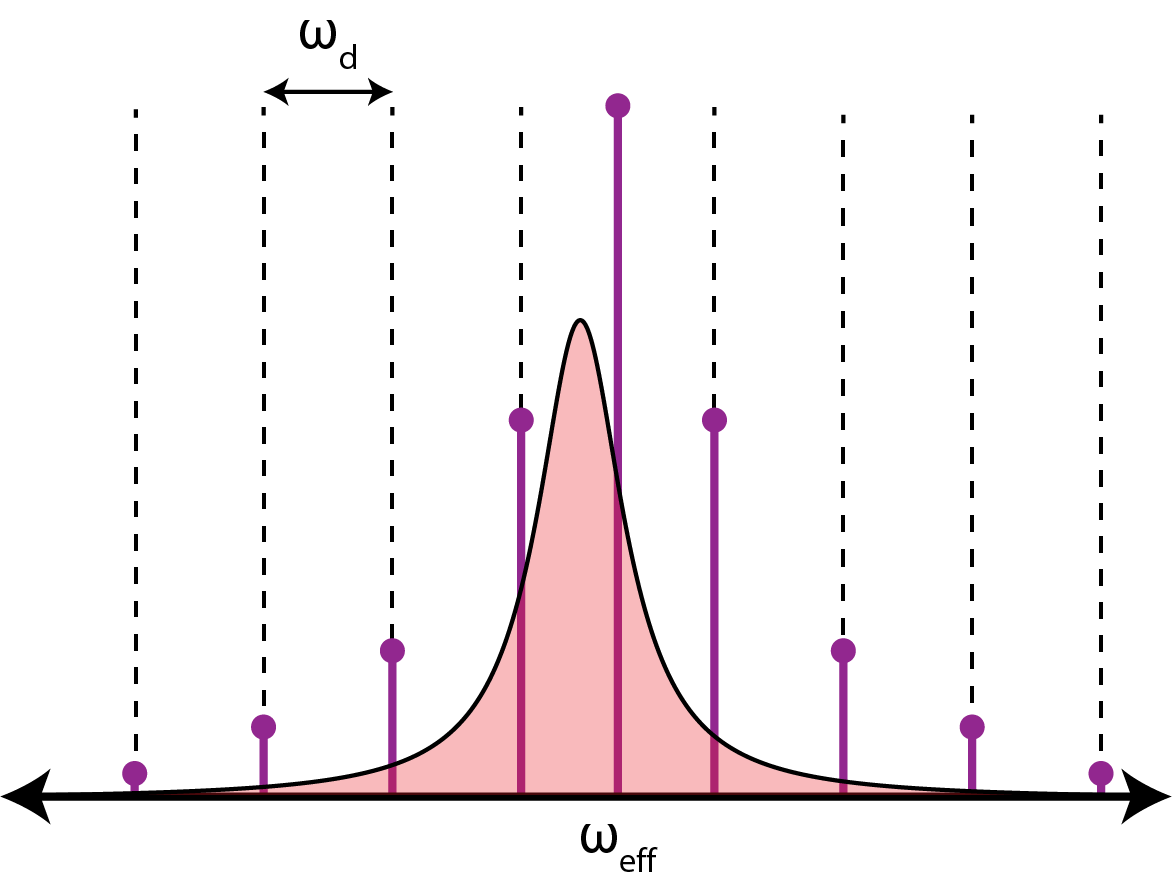}
    \caption{Schematic showing the overlap between the electron noise spectral density (red peak) and the filter function of the Floquet driving protocol (purple sticks). The spectral comb is centered around the effective energy gap of the nuclear spins in the interaction frame $\omega_{\rm eff}$, and is spaced at integer intervals of the driving frequency, $\omega_{d}$, with the amplitude of each stick given by the Fourier coefficients $c_{k}$.}
    \zfl{spectral_overlap}
\end{figure}

\subsection{Derivation of hopping rates}
\label{subsection_hopping}

The derivation of $W^{\rm LF}$ proceeds as follows. For any two nuclear spins $i$ and $j$, we can write the truncated Hamiltonian describing the flip-flop processes as
\begin{equation*}
H_{ij} + H_{1}(t)
\end{equation*}
where $H_{ij}$ describes energy-conserving flip-flops between spins $i$ and $j$, and $H_{1}(t)$ describes time-dependent local fields from the surrounding bath spins. While the relatively low magnetic field ($B_{0}=36$ mT) can permit coupling of the nuclear Zeeman reservoir to excitations of the electronic dipolar reservoir, a direct contribution to $^{13}$C-$^{13}$C polarization exchange would require higher-order many-body processes, for example correlated transitions involving multiple electron spins and multiple nuclear spins.  Such effects are neglected in the present work. Instead, both the nuclear and electron baths are treated as sources of stochastic longitudinal field fluctuations, which captures the dominant broadening of the nuclear zero-quantum spectrum and the resulting spin-transport dynamics. $H_{ij}$ is given by
\begin{equation*}
H_{ij} = -\frac{1}{4}d_{ij}(I^{+}_{i}I^{-}_{j} + I^{-}_{i}I^{+}_{j}),
\end{equation*}
and $H_{1}(t)$ is given by 
\begin{equation*}
H_{1}(t) = H_{n}(t) + H_{e}(t)
\end{equation*}
where $H_{n}(t)$ and $H_{e}(t)$ describe the time-dependent local fields originating from the nuclear and electronic spin baths, respectively, and can be written as follows:
\begin{equation*}
H_{n}(t) = \sum_{k\neq i,j}d_{ik}I^{z}_{i}I^{z}_{k}(t) + d_{jk}I^{z}_{j}I^{z}_{k}(t)
\end{equation*}
and 
\begin{equation*}
H_{e}(t) = \sum_{\mu}d_{i\mu}I^{z}_{i}S^{z}_{\mu}(t) + d_{j\mu}I^{z}_{j}S^{z}_{\mu}(t)
\end{equation*}
where $\mu$ represents paramagnetic impurities (NV and P1 centers). Equivalently, one can write:
\begin{equation*}
H_{1}(t) = I^{z}_{i}B_{i}(t) + I^{z}_{j}B_{j}(t)
\end{equation*}
with
\begin{equation*}
B_{i}(t) = \sum_{k\neq i,j}d_{ik}I^{z}_{k}(t) + \sum_{\mu}d_{i\mu}S^{z}_{\mu}(t)
\end{equation*}
and similarly for $B_{j}(t)$. Now, transforming into the interaction frame of $H_{1}(t)$, $I^{\pm}_{i}$ transforms as 
\begin{equation*}
I^{\pm}_{i} \rightarrow I^{\pm}_{i}e^{\pm i\int_{0}^{t}B_{i}(t')dt'}
\end{equation*}
and similarly for $I^{\pm}_{j}$. Hence, $H_{ij}$ becomes 
\begin{equation*}
\tilde{H}_{ij} = -\frac{1}{4}d_{ij}[I^{+}_{i}I^{-}_{j}e^{i\phi_{ij}(t)} + I^{-}_{i}I^{+}_{j}e^{-i\phi_{ij}(t)}]
\end{equation*}
where
\begin{equation*}
\phi_{ij}(t) = \int_{0}^{t} \Omega_{ij}(t') dt'
\qquad \text{and} \qquad
\Omega_{ij}(t) = B_{i}(t) - B_{j}(t).
\end{equation*}
Applying second order perturbation theory, the probability of a transition from state $\ket{\downarrow_{i}\uparrow_{j}}$ to state $\ket{\uparrow_{i}\downarrow_{j}}$ $P_{ij}$ is then given by:
\begin{equation*}
P_{ij} = \left[-\frac{1}{4}d_{ij}\right]^{2} \int_{0}^{t}dt_{1}\int_{0}^{t}dt_{2} \langle e^{-i\phi_{ij}(t_{1})} e^{+i\phi_{ij}(t_{2})} \rangle
\end{equation*}
and similarly for $\ket{\uparrow_{i}\downarrow_{j}}$ to $\ket{\downarrow_{i}\uparrow_{j}}$~\cite{slichterPrinciplesMagneticResonance1990}. Assuming the noise is stationary, in the long-time limit (i.e., when $t \gg \tau_{c}$ where $\tau_{c}$ is the bath correlation time) the transition rate $W_{ij}$ can be written as
\begin{equation*}
W_{ij} = \left[-\frac{1}{4}d_{ij}\right]^{2} \int_{-\infty}^{\infty} \langle e^{-i[\phi_{ij}(\tau) - \phi_{ij}(0)]} \rangle d\tau,
\end{equation*}
where $\tau = t_{1} - t_{2}$. To evaluate the integral, we assume the noise is Gaussian with zero mean. This yields
\begin{equation*}
W_{ij} = \left[-\frac{1}{4}d_{ij}\right]^{2} \int_{-\infty}^{\infty} e^{-\frac{1}{2} \langle \phi_{ij}^{2}(|\tau|) \rangle} d\tau.
\end{equation*}
Evaluating $\langle \phi_{ij}^{2}(t) \rangle$ yields,
\begin{equation*}
\langle \phi_{ij}^{2}(t) \rangle = \bigg\langle \int_{0}^{t} \Omega_{ij}(t_{1}) dt_{1} \int_{0}^{t} \Omega_{ij}(t_{2}) dt_{2} \bigg\rangle = \int_{0}^{t} dt_{1} \int_{0}^{t} dt_{2} \langle \Omega_{ij}(t_{1})\Omega_{ij}(t_{2}) \rangle.
\end{equation*}
Again, assuming the stationary noise, in the long-time limit we get
\begin{equation*}
\langle \phi_{ij}^{2}(t) \rangle = 2t\int_{0}^{\infty} \langle \Omega_{ij}(\tau)\Omega_{ij}(0) \rangle d\tau.
\end{equation*}
Defining $\Gamma_{ij}$ as
\begin{equation*}
\Gamma_{ij} = \int_{0}^{\infty} \langle \Omega_{ij}(\tau)\Omega_{ij}(0) \rangle d\tau,
\end{equation*}
$W_{ij}$ becomes 
\begin{equation*}
W_{ij} = \left[-\frac{1}{4}d_{ij}\right]^{2} \int_{-\infty}^{\infty} e^{-\Gamma_{ij} |\tau|} d\tau
\end{equation*}
which can be simplified to
\begin{equation*}
W_{ij} = \left[-\frac{1}{4}d_{ij}\right]^{2} \frac{2}{\Gamma_{ij}}.
\end{equation*}
Now, we must evaluate $\Gamma_{ij}$. We can write it as 
\begin{equation*}
\Gamma_{ij} = \int_{0}^{\infty} \Bigg\langle \left[ \sum_{k\neq i,j} (d_{ik}-d_{jk})I^{z}_{k}(t) + \sum_{\mu} (d_{i\mu}-d_{j\mu})S^{z}_{\mu}(t) \right] \left[ \sum_{k'\neq i,j} (d_{ik'}-d_{jk'})I^{z}_{k'}(0) + \sum_{\mu'} (d_{i\mu'}-d_{j\mu'})S^{z}_{\mu'}(0) \right] \Bigg\rangle dt.
\end{equation*}
Ignoring cross terms (i.e., assuming the bath spins are uncorrelated with each other), this can be simplified to
\begin{equation*}
\Gamma_{ij} = g_{n}\sum_{k\neq i,j} (d_{ik}-d_{jk})^{2} + g_{e}\sum_{\mu} (d_{i\mu}-d_{j\mu})^{2} 
\end{equation*}
where $g_{n} = \int_{0}^{\infty}\langle I^{z}(t)I^{z}(0) \rangle dt \
 \text{and} \ g_{e} = \int_{0}^{\infty}\langle S^{z}(t)S^{z}(0) \rangle dt$
and we have assumed the correlation function is the same for all nuclear spins and all electron spins, respectively. We further assume that the nuclear and electron bath autocorrelation functions decay exponentially with characteristic correlation times $\tau_n$ and $\tau_e$, respectively. For spin-$1/2$ particles, this yields $g_{n} = \tau_{n}/4$ and $g_{e} = \tau_{e}/4$.
Thus, finally, the transition rate is given by 
\begin{equation*}
W_{ij} = \frac{d_{ij}^{2}/2}{\tau_{n}\sum_{k\neq i,j} (d_{ik}-d_{jk})^{2} + \tau_{e}\sum_{\mu} (d_{i\mu}-d_{j\mu})^{2}}
\end{equation*}

To estimate the nuclear and electronic bath correlation times $\tau_{n}$ and $\tau_{e}$, we compute the second moments of the nuclear-nuclear and electron-electron dipolar coupling distributions, $M_{2,n}$ and $M_{2,e}$, respectively. For a given disorder realization, we compute the second moments as
\begin{equation*}
M_{2,n}=\Big\langle \sum_{j\neq i} d_{ij}^{2}\Big\rangle_i
\qquad \text{and} \qquad
M_{2,e}=\Big\langle \sum_{\nu\neq \mu} d_{\mu\nu}^{2}\Big\rangle_{\mu}.
\end{equation*}
The associated correlation times are then estimated using the Gaussian free-induction-decay relation~\cite{abragamPrinciplesNuclearMagnetism1983a}
\begin{equation*}
\tau_{n} = \sqrt{\frac{\pi}{8}}\frac{1}{\sqrt{M_{2,n}}}
\qquad \text{and} \qquad
\tau_{e} = \sqrt{\frac{\pi}{8}}\frac{1}{\sqrt{M_{2,e}}}.
\end{equation*}
To ensure convergence with respect to simulation volume, the calculation was repeated for increasing lattice sizes and averaged over 1000 disorder realizations for each size. The resulting values of $\tau_{n}$ and $\tau_{e}$ approach a well-defined plateau as the number of simulated spins increases. The converged correlation times obtained from this procedure are $\tau_{e} = 19$ ns and $\tau_n = 6.67$ ms, which are used throughout the simulations.

\section{Kullback--Leibler divergence}
\label{section_kl}
To quantify the distance of a nonequilibrium nuclear-spin state from equilibrium, we employ the Kullback--Leibler (KL) divergence, or relative entropy. The KL divergence measures the distinguishability between two probability distributions and has been shown to provide a rigorous metric for identifying the Mpemba effect~\cite{luNonequilibriumThermodynamicsMarkovian2017, aresQuantumMpembaEffects2025}. In the present case, the two distributions correspond to the nonequilibrium polarization distribution at time $t$, denoted by $P(t)$, and the equilibrium distribution, denoted by $Q$.

We define the polarization of nuclear spin $i$ as
\begin{equation*}
p_i(t)=\langle 2I_i^z\rangle,
\end{equation*}
such that $p_i(t)\in[-1,1]$. The probabilities of finding spin $i$ in the states $\ket{\uparrow}$ and $\ket{\downarrow}$ are therefore
\begin{equation*}
P_{i,\uparrow}(t)=\frac{1+p_i(t)}{2},
\qquad
P_{i,\downarrow}(t)=\frac{1-p_i(t)}{2}.
\end{equation*}

In the rotating frame, the nuclear-spin ensemble relaxes toward an effectively infinite-temperature state with vanishing equilibrium polarization, $p_{\rm eq}=0$. Consequently, the equilibrium probability distribution for each spin is
\begin{equation*}
Q_{i,\uparrow}=Q_{i,\downarrow}=\frac12.
\end{equation*}

Let
\begin{equation*}
\vec{\sigma}=(\sigma_1,\sigma_2,\dots,\sigma_N),
\qquad
\sigma_i\in\{\ket{\uparrow},\ket{\downarrow}\},
\end{equation*}
denote a many-body spin configuration. For notational simplicity, we temporarily suppress the explicit time dependence of $P$. To construct a tractable measure of the distance to equilibrium, we neglect inter-spin correlations and approximate the many-body probability distribution as a product of single-spin probability distributions,
\begin{equation*}
P(\vec{\sigma})=\prod_{i=1}^N P_i(\sigma_i),
\qquad
Q(\vec{\sigma})=\prod_{i=1}^N Q_i(\sigma_i).
\end{equation*}

The KL divergence between the nonequilibrium and equilibrium distributions is then defined as
\begin{equation*}
D_{\rm KL}(P|Q)
=
\sum_{\vec{\sigma}}
P(\vec{\sigma})
\log_2
\left(
\frac{P(\vec{\sigma})}{Q(\vec{\sigma})}
\right).
\end{equation*}

Under the product-state approximation, the relative entropy becomes additive,
\begin{equation*}
D_{\rm KL}(P|Q)
=
\sum_{i=1}^N
D_{\rm KL}(P_i|Q_i),
\end{equation*}
which yields
\begin{equation*}
D_{\rm KL}(P|Q)
=
\sum_{i=1}^N
\left[
\frac{1+p_i(t)}{2}
\log_2\!\left(1+p_i(t)\right)
+
\frac{1-p_i(t)}{2}
\log_2\!\left(1-p_i(t)\right)
\right].
\end{equation*}